\documentclass[aps,pra,amsmath,amssymb,floatfix,twocolumn,groupedaddress,nofootinbib,10pt]{revtex4-1}
 
\usepackage{times}
\usepackage{bbold}
\usepackage{bm}
\usepackage{graphicx} 
\usepackage{hyperref}
\usepackage{textcomp}
\usepackage{amssymb} 
\usepackage{wasysym}
\usepackage{epstopdf}
\usepackage{comment}

\newcommand{\be}{\begin{eqnarray}}
\newcommand{\ee}{\end{eqnarray}}

\newcommand{\app}{Appendix~\ref{Appendix:Details}}

\newcommand{\TLL}{{Tomonaga-Luttinger liquid}}

\newcommand{\kFj}{k_{ Fj}}
\newcommand{\kFo}{k_{ F1}}
\newcommand{\kFt}{k_{ F2}}

\newcommand{\da}{^{\dagger}}
\newcommand{\hc}{{\rm H.c.}}

\newcommand{\Hov}{H_{\rm B}}
\newcommand{\Bov}{B}
\newcommand{\Oov}{O_B}

\newcommand{\Rjs}{R_{j\sigma} }
\newcommand{\Ljs}{L_{j\sigma} }

\newcommand{\Rou}{R_{1\uparrow} }
\newcommand{\Rod}{R_{1\downarrow} }

\newcommand{\Lou}{L_{1\uparrow} }
\newcommand{\Lod}{L_{1\downarrow} }

\newcommand{\Rtu}{R_{2\uparrow} }
\newcommand{\Rtd}{R_{2\downarrow} }

\newcommand{\Ltu}{L_{2\uparrow} }
\newcommand{\Ltd}{L_{2\downarrow} }

\newcommand{\phiRtu}{\phi_{R {\rm 2\uparrow}} }
\newcommand{\phiRtd}{\phi_{R {\rm 2\downarrow}} }

\newcommand{\phiLtu}{\phi_{L {\rm 2\uparrow}} }
\newcommand{\phiLtd}{\phi_{L {\rm 2\downarrow}} }

\begin{document}
 
\title{Universal conductance dips and fractional excitations in a two-subband quantum wire}
 
\author{Chen-Hsuan Hsu$^{1}$} 
\author{Flavio Ronetti$^{2}$}
\author{Peter Stano$^{1,3,4}$}
\author{Jelena Klinovaja$^{2}$}
\author{Daniel Loss$^{1,2}$} 

\affiliation{$^{1}$RIKEN Center for Emergent Matter Science (CEMS), Wako, Saitama 351-0198, Japan}
\affiliation{$^{2}$Department of Physics, University of Basel, Klingelbergstrasse 82, CH-4056 Basel, Switzerland} 
\affiliation{$^{3}$Department of Applied Physics, School of Engineering, University of Tokyo, 7-3-1 Hongo, Bunkyo-ku, Tokyo 113-8656, Japan}
\affiliation{$^{4}$Institute of Physics, Slovak Academy of Sciences, 845 11 Bratislava, Slovakia}

\date{\today}
                                                     
\begin{abstract}
We theoretically investigate a quasi-one-dimensional quantum wire, where the lowest two subbands are populated, in the presence of a helical magnetic field. We uncover a backscattering mechanism involving the helical magnetic field and Coulomb interaction between the electrons. The combination of these ingredients results in scattering resonances and partial gaps which give rise to nonstandard plateaus and conductance dips at certain electron densities. The positions and values of these dips are independent of material parameters, serving as direct transport signatures of this mechanism.
Our theory applies to generic quasi-one-dimensional systems, including a Kondo lattice and a quantum wire subject to intrinsic or extrinsic spin-orbit coupling.
Observation of the universal conductance dips would identify a strongly correlated fermion system hosting fractional excitations, resembling the fractional quantum Hall states.  
\end{abstract}

\maketitle
 
\section{Introduction}
Quasi-one-dimensional conductors, such as semiconducting nanowires or quantum point contacts, are typical elements of nanocircuits. On the one hand, they embody the ultimate quantum limit upon shrinking a conductor. On the other hand, they provide a testbed for fundamental physics of low-dimensional interacting fermions. From both aspects, conductance is the quantity of prime interest and  of most direct experimental access. The observation of conductance quantization at integer multiples of the conductance quantum $G_0= e^2/h$, where $e$ is the elementary charge and $h$ is the Planck constant, was a landmark achievement~\cite{vanWees:1988,Wharam:1988} in experimental realization of conductors in the quantum limit. It initiated extensive research activities on the quantum conductance, both in experiment and theory, which continue unabated.
While the ballistic conductance is expected to be robust against interactions~\cite{Maslov:1995,Ponomarenko:1995,Safi:1995}, deviations from the universal values are routinely observed~\cite{Tarucha:1995,Yacoby:1996,Thomas:1996,Cronenwett:2002,Micolich:2011,Bauer:2013,Iqbal:2013,Scheller:2014,Heedt:2017,Kammhuber:2017,EstradaSaldana:2018,Kumar:2019,Mittag:2019}, including mysterious conductance features that are unexpected from standard single-particle quantum mechanics, such as dips and new plateaus at fractional conductance values, strongly suggesting the importance of many-body interaction effects~\cite{Maslov:1995b,Meir:2002,Rejec:2006,Pedder:2016,Schimmel:2017,Aseev:2017}.

In addition to hosting strongly interacting fermions, the quasi-one-dimensional systems are capable of stabilizing topological phases upon combining with an external magnetic field, spin-orbit coupling, and superconductivity~\cite{Lutchyn:2010,Oreg:2010,Klinovaja:2012a}. 
The observation of zero-bias conductance peaks in proximitized Rashba nanowires~\cite{Mourik:2012,Das:2012,Deng:2012}, which hint at the presence of Majorana bound states, has stimulated numerous sequential studies on topological aspects of the quasi-one-dimensional systems~\cite{Prada:2019}. It motivated alternative setups for the realization of Majorana bound states, in which the external magnetic field and spin-orbit coupling are replaced by other ingredients. One example is to replace these ingredients with a helical magnetic field, which arises from either intrinsic spin textures~\cite{Klinovaja:2013,Braunecker:2013,Vazifeh:2013,Nadj-Perge:2013,Nadj-Perge:2014,Pientka:2014,Hsu:2015,Pawlak:2016} or artificially synthesized nanomagnets~\cite{Klinovaja:2012b,Desjardins:2019}. Alternatively, topological phases can also be stabilized by sufficiently strong electron-electron interactions in a quasi-one-dimensional geometry with multiple conducting channels, such as nanowires, edges of a two-dimensional topological insulator, and hinges of a higher-order topological insulator~\cite{Gaidamauskas:2014,Klinovaja:2014a,Klinovaja:2014b,Klinovaja:2014c,Klinovaja:2015,Ebisu:2016,Schrade:2017,Thakurathi:2018,Hsu:2018b,Haim:2019}. Interestingly, incorporating these ingredients has led to the discovery of not only new candidate platforms but also more exotic topological phases characterized by, e.g., parafermions or fractionally charged fermions.  
   
Here we merge these ingredients by considering an interacting two-subband quantum wire in the presence of a helical (spatially rotating) magnetic field, which induces a spin-selective partial gap in the lower subband.
The helical field can be generated in a wide variety of systems, including a Kondo lattice with localized spins ordered by the Ruderman-Kittel-Kasuya-Yosida (RKKY) interaction~\cite{Braunecker:2009a,Braunecker:2009b,Klinovaja:2013,Meng:2013,Stano:2014,Meng:2014a,Hsu:2015,Hsu:2017,Hsu:2018}, a spin-orbit-coupled wire in the presence of an external magnetic field~\cite{Braunecker:2010,Rainis:2014}, and a wire subject to spatially modulated nanomagnets~\cite{Klinovaja:2012b,Desjardins:2019}. 
While it is well known that such a helical field, being equivalent to the combination of Zeeman and spin-orbit fields~\cite{Braunecker:2010}, is crucial for the nontrivial topology in the paradigm setup based on a single-subband Rashba wire, there is less attention on its role in a wire where higher subbands are populated.
 
Remarkably, in this setting we identify a mechanism leading to fractional conductance quantization. It is manifested by dips on the second conductance plateau of a quantum wire, appearing when the Fermi wave vectors of the two lowest subbands become commensurate.
The effect arises in the presence of the helical magnetic field and strong electron-electron interactions. The resulting higher-order backscattering opens a gap in the upper subband, leading to {\it universal conductance dips}---their positions and values are independent of material details. 
Interestingly, upon varying the carrier density and/or the wire width through voltage gates, the system reveals the formation of strongly correlated fermions at certain gate voltages, where the partially gapped upper subband forms a fractional \TLL~\cite{Kane:2002,Laughlin:1983}, in analogy to the fractional quantum Hall edge states.

The paper is organized as follows. In Sec.~\ref{Sec:Setup} we describe two types of setups that realize a helical magnetic field and explain how to achieve the commensurability condition in each of them.  
In Sec.~\ref{Sec:Helix} we explain how the helical magnetic field induces a gap opening in the electron energy spectrum. 
The direct transport signatures of our theory, universal conductance dips on the second plateau, and the conditions for their appearance are presented in Sec.~\ref{Sec:Dip}. The higher-order scatterings for the even- and odd-denominator fillings are discussed in Sec.~\ref{Sec:Even} and Sec.~\ref{Sec:Odd}, respectively. Finally, we discuss the experimental realization and verification of our theory in Sec.~\ref{Sec:Discussion}. 
The details of the calculation are given in Appendix~\ref{Appendix:Details}. The estimation for the interaction parameter is presented in Appendix~\ref{Appendix:Kjc}.

\begin{figure}[t]
\includegraphics[width=\linewidth]{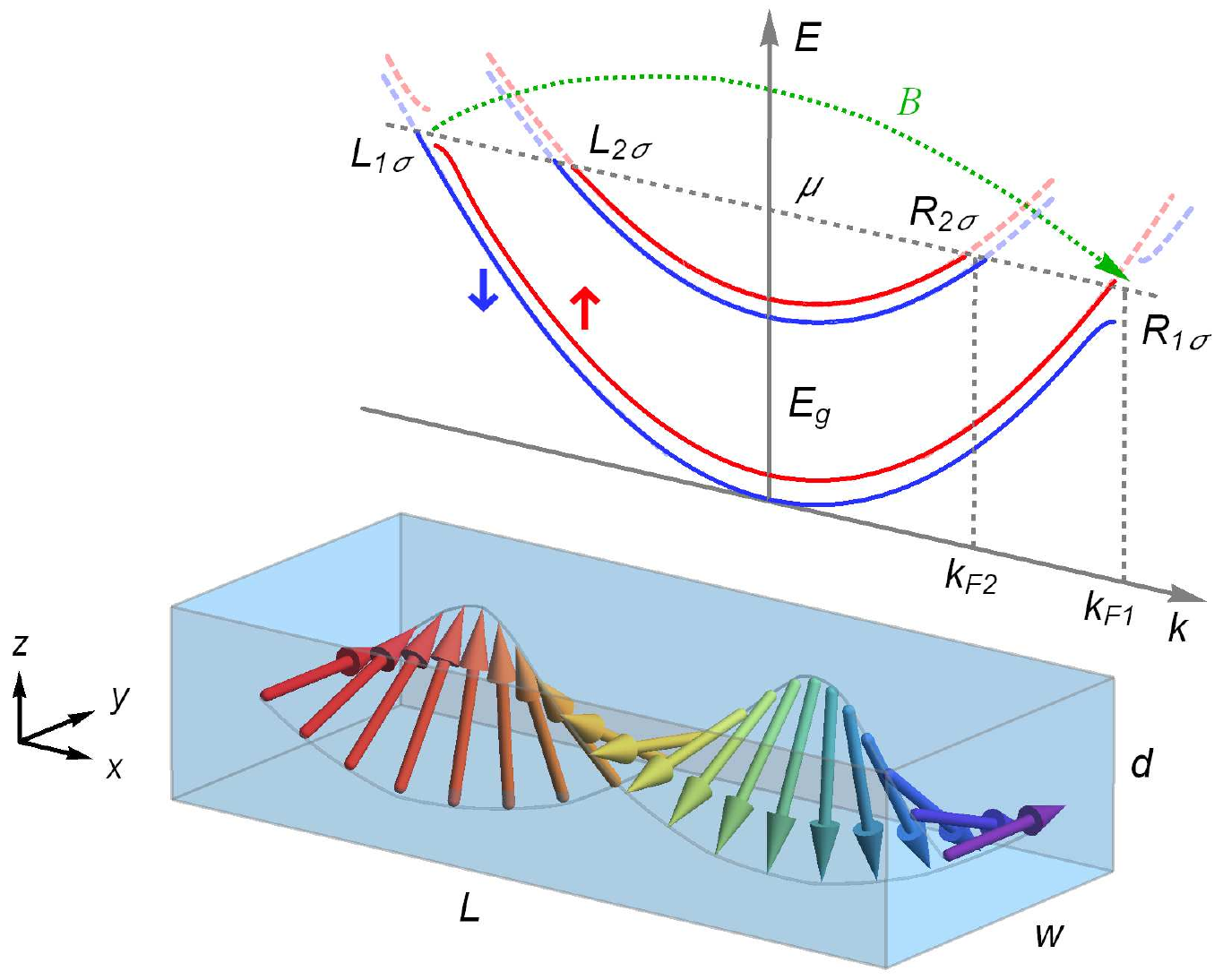}
\caption{Spectrum of a wire with two transverse subbands, each being spin-degenerate; for clarity, the two spin branches are separated and plotted in different colors. 
The ratio of the chemical potential $\mu$ to the subband spacing $E_g$ is adjusted to fulfill the commensurability condition given in Eq.~\eqref{Eq:condition}.
The helical magnetic field ${\bf \Bov}$ leads to $2\kFo$ spin-flip backscattering (green arrow) [see Eq.~\eqref{Eq:g1}] and a partial gap in the lower subband.
In a wire (blue block) with anisotropic confinements, $d < w \ll L$, electrons experience a spatially rotating magnetic field (colored arrows) described by Eq.~\eqref{Eq:Bov}.
}
\label{Fig:Setup}
\end{figure}

\section{Setup~\label{Sec:Setup}}
We consider a quantum wire with anisotropic transverse confinements. The geometry is chosen to separate the subbands corresponding to the stronger confinement direction so that the chemical potential, tuned by a voltage gate, intersects with the lowest two transverse subbands, each being spin-degenerate (see Fig.~\ref{Fig:Setup}).
Our mechanism applies to any wires with well-defined subbands, irrespective of the specific form of the transverse confinement potential.
In this two-subband regime, the electron operator can be expanded around the Fermi points $\pm \kFj$,
\begin{align}
\psi_{j\sigma} (x) =& e^{i \kFj x} \Rjs (x) + e^{-i \kFj x} \Ljs (x) , 
\label{Eq:SlowVarying}
\end{align}
with the slowly varying right(left)-moving fields $\Rjs$ ($\Ljs$), the subband index $j \in \{ 1,2\}$, and the spin index $\sigma \in \{ \uparrow , \downarrow \}$.
We will suppress the coordinate $x$ along the wire in the argument unless it may cause confusion.
In analogy to the fractional quantum Hall states, we define the ratio $\nu \equiv \kFt / \kFo$ as the ``filling factor''.
The commensurability condition is fulfilled when its inverse, $1/\nu$, is an integer. In terms of the chemical potential measured from the bottom of the lowest subband, the condition reads 
\be
\left.\frac{\mu}{E_g}\right|_{\nu} &=& \frac{1}{1-\nu^2},
\label{Eq:condition}
\ee
with the spacing $E_g$ between the two subbands; see Fig.~\ref{Fig:Setup}.

We are interested in the setting where the electrons experience a helical magnetic field in the following form,
\be
{\bf \Bov}(x) &=& \Bov \big[ {\bf e}_y \cos (2\kFo x)  + {\bf e}_z \sin (2\kFo x) \big],
\label{Eq:Bov}
\ee
with the unit vector ${\bf e}_{\mu}$ and field strength $\Bov$. Without loss of generality, we assign the up/down-spin orientation to be in the $\pm x$ direction and choose the clockwise rotating magnetic field. 
The spatial pitch $\pi/\kFo$ of the field is chosen so that it causes spin-flip backscattering between right- and left-moving electrons in the lower subband. 
We remark that it is sufficient to have a helical field with the dominant Fourier component at $2\kFo$ even without a perfect periodicity~\cite{Klinovaja:2013x}.
To achieve Eqs.~\eqref{Eq:condition} and \eqref{Eq:Bov} simultaneously, we consider the following two types of setups.

\begin{figure}[t]
\centering
\includegraphics[width=0.95\linewidth]{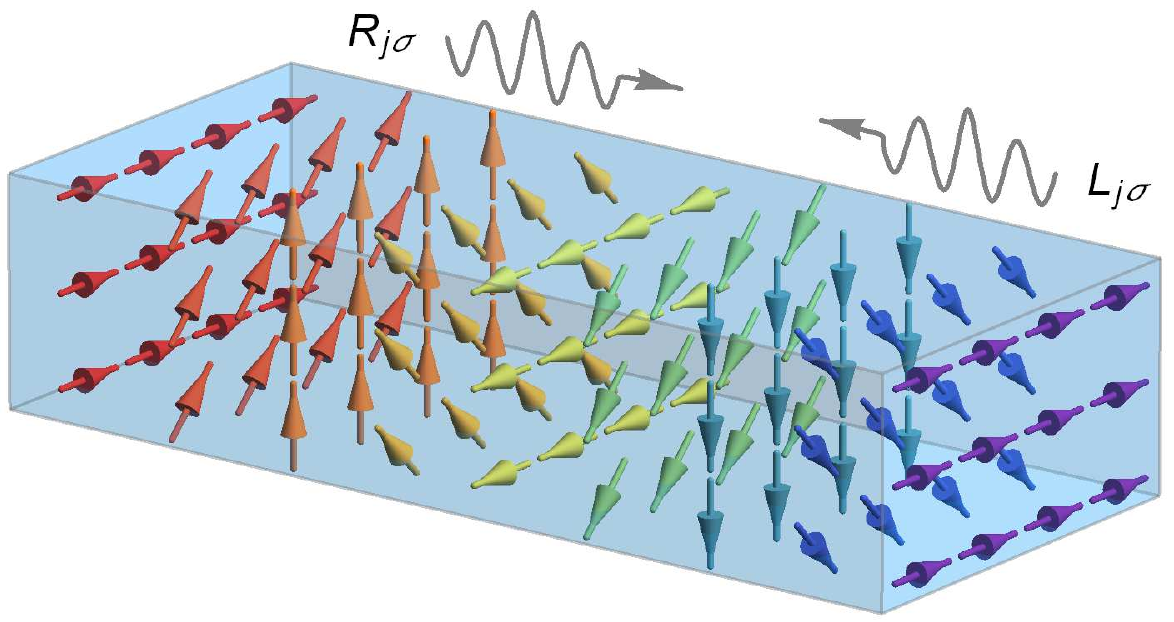}
\caption{A quasi-one-dimensional Kondo lattice is realized in a wire (blue block) hosting localized spins (arrows). The itinerant charge carriers (labeled by $R_{j\sigma}$ and $L_{j\sigma}$) mediate the RKKY coupling and lead to the formation of a spin helix at low temperature, where the localized spins align ferromagnetically within a cross section and form a helical pattern along the wire. 
For clarity, localized spins in different cross sections are colored differently. 
}
\label{Fig:Setup_KL}
\end{figure}

\subsection{Self-tuning helix in a Kondo lattice~\label{SubSec:Kondo}}
In the first type, one can exploit a quasi-one-dimensional Kondo lattice, where localized spins (e.g., nuclear spins or magnetic dopants) are present and coupled to conduction electrons in the wire. The conduction electrons mediate indirect RKKY coupling between the localized spins, which can be strongly enhanced by electron-electron interactions in low dimension.
As has been established in the literature on wires where only the lowest transverse subband is populated (one-subband regime)~\cite{Braunecker:2009a,Braunecker:2009b,Klinovaja:2013,Stano:2014,Meng:2014a,Hsu:2015,Hsu:2017,Hsu:2018}, 
the RKKY coupling stabilizes a helical spin order in a finite-length wire at sufficiently low temperature; see Fig.~\ref{Fig:Setup_KL} for illustration of the helical spin order in a Kondo lattice.
The helically ordered spins induce a spatially rotating magnetic field, which acts back on the electrons. 
Since the RKKY coupling arises from resonant scattering of electrons at the Fermi energy, the spatial period of the helix and thus the helical field is determined by the Fermi wave vector $\kFo$ itself. As a consequence, the spatial period of the helical field is self-tuned to $\pi/\kFo$  for any chemical potential in the one-subband regime. 
 
When the chemical potential is adjusted to populate the second transverse subband (two-subband regime), the $\pi/\kFo$ helix remains stabilized by the lower-subband electrons. A second, additional helix with a spatial pitch of $\pi/\kFt$ could be induced by the upper-subband electrons~\cite{Meng:2013}. 
However, since the two helices have in general different ordering temperatures, it is possible, by adjusting the temperature, to reach the regime in which only the $\pi/\kFo$ helix is present. 
As outlined here, previous work focused on either the one-subband regime or the double-helix phase in the two-subband regime with incommensurable configuration between $\kFo$ and $\kFt$ (that is, the ratio $\kFo / \kFt$ is not an integer). In contrast, we turn our attention to the temperature range where there is a single helix, which induces a helical magnetic field of the form in Eq.~\eqref{Eq:Bov} when the chemical potential is in the two-subband regime. In this Kondo-lattice setup, the commensurability condition in Eq.~\eqref{Eq:condition} can be achieved by scanning the chemical potential $\mu$ via a voltage gate in a fixed wire geometry (hence a fixed subband spacing $E_g$).

\subsection{Artificial helical magnetic field~\label{SubSec:Alternative}}
\begin{figure}[t]
\centering
\includegraphics[width=0.9\linewidth]{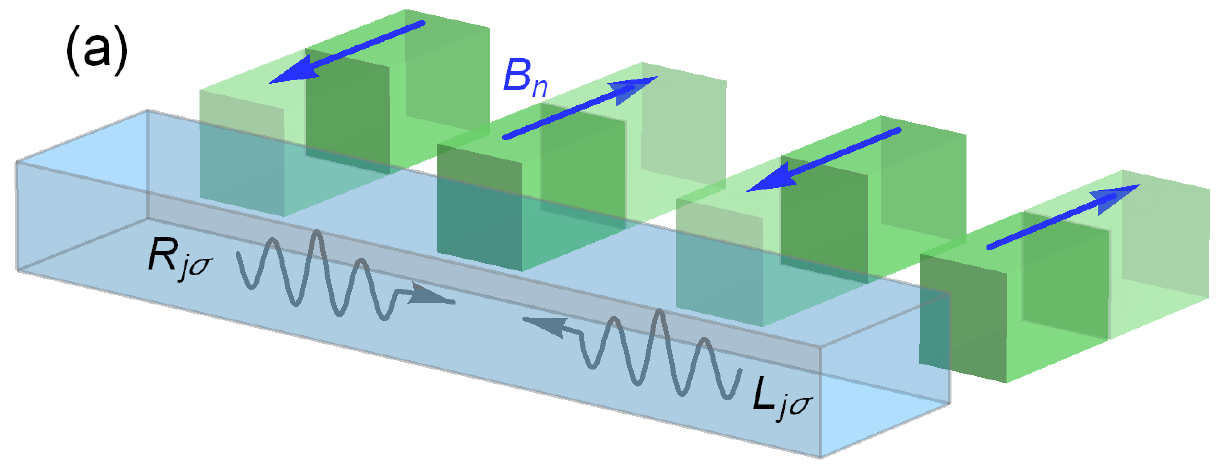} \\
\vspace{3pt}
\includegraphics[width=0.33\linewidth]{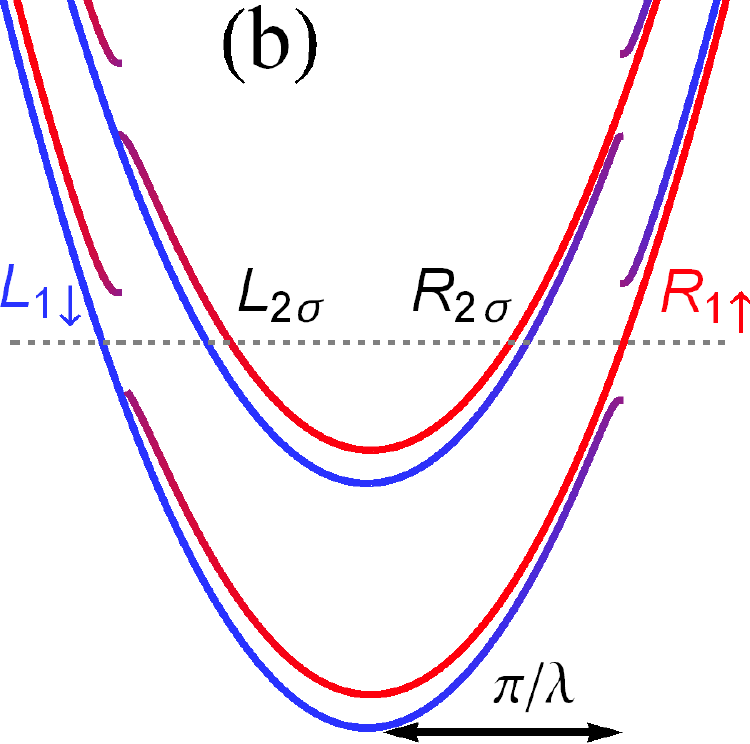}
\includegraphics[width=0.66\linewidth]{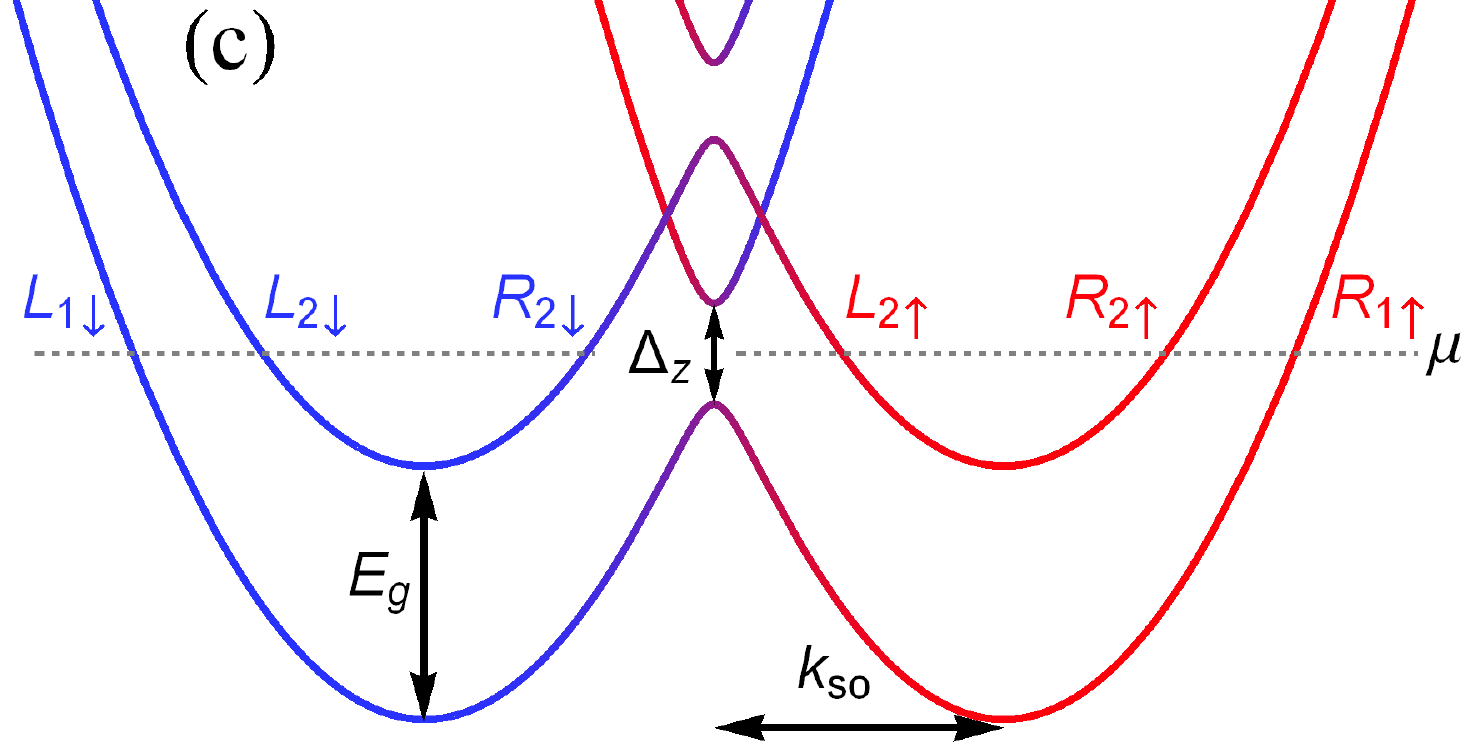} 
\caption{Setups realizing an artificial helical magnetic field.
(a) By depositing nanomagnets with alternative magnetization (green blocks) in proximity to a quantum wire (blue block), a spatially varying magnetic field ${\bf B_n}(x)$ is generated, coupling to the conduction electrons (labeled by $R_{j\sigma}$ and $L_{j\sigma}$) in the wire. 
It leads to a partial gap at finite momentum in the spectrum, as shown in Panel (b); for clarity, the two spin branches are shifted vertically and plotted in different colors.
(c) In an alternative setup based on a Rashba wire, the combination of the Zeeman and spin-orbit fields leads to a partial gap at zero momentum.  
The energy spectra in Panels~(b) and (c) can be mapped onto each other through a gauge transformation (see the text).
When the chemical potential is tuned into the partial gap, the required helical magnetic field for our mechanism is realized.
}
\label{Fig:Setup_NM}
\end{figure}

As an alternative to the Kondo-lattice setup, one can generate the helical magnetic field in artificially engineered nanostructures. 
For concreteness, here we provide two examples. 
In the first example, spatially modulated nanomagnets are deposited in proximity to a wire~\cite{Klinovaja:2012b,Desjardins:2019}. As illustrated in Fig.~\ref{Fig:Setup_NM}(a), the nanomagnets induce a spatially rotating magnetic field ${\bf B_n}(x)$ with a spatial pitch of $\lambda$, leading to a partial gap opening at finite momentum $\pm \pi/\lambda$ [see Fig.~\ref{Fig:Setup_NM}(b)]. The second example is based on a Rashba wire, in which the spin-orbit coupling causes a spin-dependent momentum shift $\pm k_{so}$ of the quadratic energy band~\cite{Braunecker:2010,Rainis:2014}. 
Applying a Zeeman field perpendicular to the spin-orbit field leads to a partial gap $\Delta_z$ opening at zero momentum, as displayed in Fig.~\ref{Fig:Setup_NM}(c). 
For $\lambda = \pi / k_{so}$, the energy spectra of the two examples can be mapped onto each other via a spin-dependent gauge transformation, which shifts the momentum $k$ by an amount of $\pm k_{so}$.

In either example, a spin-selective partial gap is opened as a result of the spin-flip backscattering between right- and left-moving electrons. 
When the chemical potential is tuned into the partial gap in the lower subband, the spectrum is equivalent to that of a wire subject to the helical magnetic field given in Eq.~\eqref{Eq:Bov}.
We note that there is an additional partial gap in the upper subband, but it is 
separated in energy from the electrons near the Fermi level and therefore does not affect the linear-response conductance in the regime of our interest.
Upon adjusting the wire width and therefore the subband spacing (for instance, through side voltage gates), one can change the ratio of $\mu/E_g$ in order to fulfill the commensurability condition given in Eq.~\eqref{Eq:condition}.

In both types (either using a Kondo lattice or artificially engineered nanostructure), a helical magnetic field is generated in a two-subband wire as illustrated in Fig.~\ref{Fig:Setup}, where the ratio of $\mu/E_g$ serves as a gate-tunable variable in conductance measurements.
Below we discuss our mechanism in a general setup and we shall give remarks on specific experimental realizations in Sec.~\ref{Sec:Discussion}.

\begin{figure*}[t]
\includegraphics[width=1.0\linewidth]{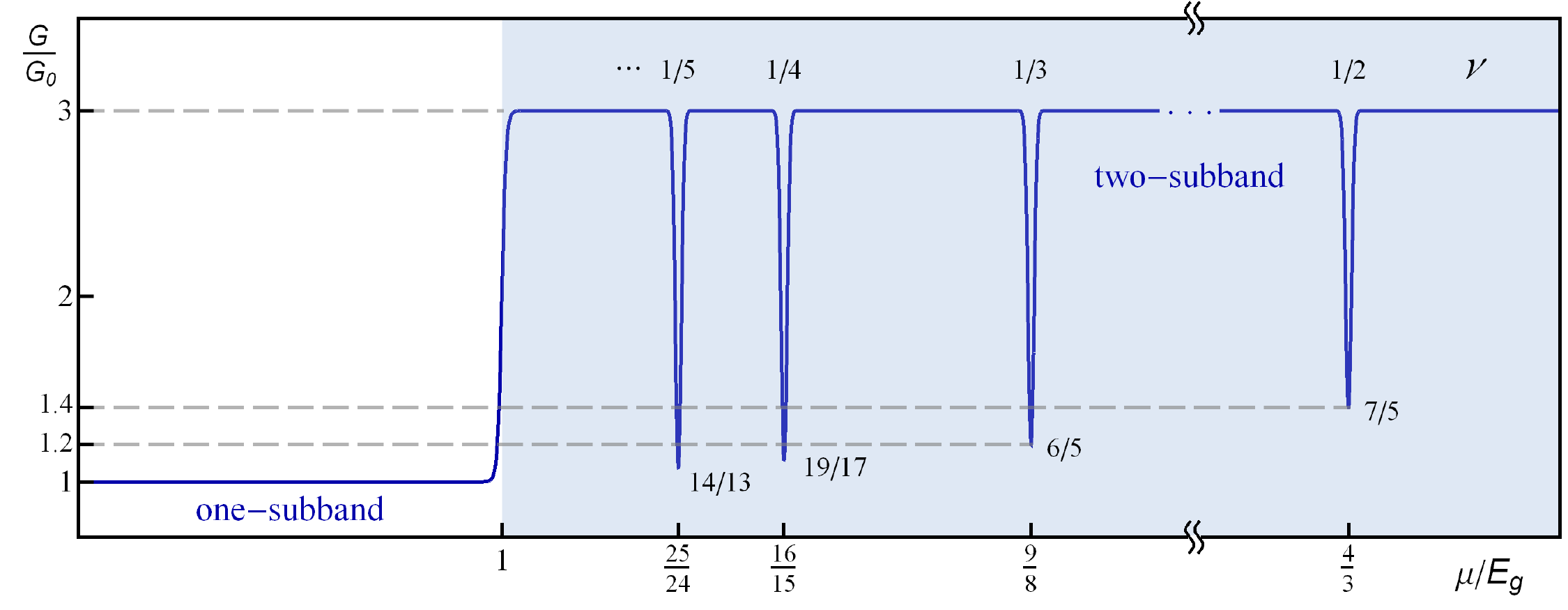}
\caption{Linear-response conductance ($G$) as a function of the chemical potential ($\mu$) to the subband spacing ($E_g$) ratio in the regime of $\mu/E_g \in\{0.9, 1.4\}$. In the one-subband regime, the first plateau shows a quantized value at $G_0 \equiv e^2/h$ due to the helical magnetic field, gapping out a half of conduction modes. In the two-subband regime (shaded region), conductance dips with the universal values [see Eq.~\eqref{Eq:conductance}] appear at commensurate positions [see Eq.~\eqref{Eq:condition}]. Away from the commensurate positions, the conductance is given by $3 G_0$ due to the half-gapped lower subband. Here we show the dips for $\nu = 1/2, \cdots, 1/5$, each modeled by a gaussian profile with a width of $\delta \mu / E_{g} = 10^{-3}$. A hyperbolic tangent function with the same width is used to smoothen the transition between the conductance plateaus. The gray lines are guidance for eyes. }
\label{Fig:conductance}
\end{figure*}

\section{Partial gap induced by the helical magnetic field~\label{Sec:Helix}}
We describe a general setup consisting of electrons in a two-subband wire and a helical magnetic field with the Hamiltonian: $H = H_0 + H_{\rm int} + \Hov$. The first term $H_0$ is the kinetic energy.
The second term $H_{\rm int} $, describing electron-electron interactions, can be separated into the forward-scattering $H_{\rm f} $ and backscattering $ H_{\rm b}$ parts.
The term $H_0 + H_{\rm f} $ describes a two-subband {\TLL} parametrized by the charge and spin interaction parameters $K_{jc}<1$ and $K_{js} \approx 1$ (for subband $j$). 
Finally, the helical field couples to the electron spin through
\begin{equation}  
\label{Eq:Hov}
\Hov = \sum_{\mu=x,y,z} \int dx \; \Bov^{\mu} \Big( \frac{1}{2} \sum_{j\sigma \sigma'} \psi_{j\sigma}^\dag  \sigma_{\sigma\sigma'}^{\mu} \psi_{j\sigma'} \Big),
\end{equation}
with $\Bov^{\mu}$ being the component of the helical field ${\bf \Bov}$ in Eq.~\eqref{Eq:Bov} and $\sigma^{\mu}$ the Pauli matrix defined as 
\be
\sigma^x= 
\left(
\begin{array}{cc}
1 & 0 \\
0 & -1 
\end{array}
\right), 
\; 
\sigma^y= 
\left(
\begin{array}{cc}
0 & 1 \\
1 & 0 
\end{array}
\right), 
\;
\sigma^z= 
\left(
\begin{array}{cc}
0 & -i \\
i & 0 
\end{array}
\right). 
\ee
In Eq.~\eqref{Eq:Hov}, there is a single resonant (that is, nonoscillatory) term in the integrand, 
\begin{equation}
\Oov  =  \frac{g^{(1)}}{2} \Rod\da  \Lou  + \hc, \label{Eq:g1}
 \end{equation}
with the coupling strength $g^{(1)} \propto \Bov$. The operator $\Oov$ describes spin-flip backscattering with momentum transfer $2\kFo$. If $\Oov$ is relevant in the renormalization-group (RG) sense, it can gap out the $\Rod$ and $\Lou$ modes, leading to a partial gap in the lower subband, as shown in Fig.~\ref{Fig:Setup}. The other modes ($\Rou$ and $\Lod$) remain gapless, resulting in a helical spin texture in the electron subsystem. 
The RG relevance of $\Oov$ can be examined from the RG flow equation,
\begin{align} \label{Eq:RG_g1}
\frac{d \tilde{g}^{(1)} }{d l } =& \frac{  3 - K_{1c}  }{2} \tilde{g}^{(1)},
\end{align}
where $l$ is the dimensionless scale and $\tilde{g}^{(1)} = \Bov a /(\hbar v_{F1})$ is the dimensionless coupling constant with the Fermi velocity $v_{F1}$ of the lower subband and the short-distance cutoff $a$. The RG flow equation shows that $\Oov$ is relevant for any repulsive interactions. From now on we focus on the regime in which $\tilde{g}^{(1)}$ flows to the strong-coupling limit and $\Oov$ gets ordered.
In this limit, operators that do not commute with $\Oov$ cannot be ordered.

The remaining, oscillating integrand in Eq.~\eqref{Eq:Hov} does not lead to a gap opening by itself due to momentum mismatch. However, as the subbands are commensurate, the combination of this oscillating term and the $H_{\rm b}$ term of the electron-electron interaction allows for higher-order scattering processes that preserve both the momentum and the spin. 
If such higher-order backscattering commutes with the operator $\Oov$, it can open a gap in the upper subband of the energy spectrum.
As a result, the primary experimental consequence of the helical-field-assisted backscattering is reduction of the conductance at certain fillings, which we present next.

\section{Universal conductance dips~\label{Sec:Dip}}
The conductance as a function of the ratio $\mu/E_g$ in the one- and two-subband regime is presented in Fig.~\ref{Fig:conductance}.
The conductance values at the dips, as well as the criterion for interaction strength, are given by 
\begin{align} \label{Eq:conductance}
\frac{G_\nu}{G_0} = & \frac{3\nu^2 + 1}{\nu^2 +1}, \;\; 
\left\{
\begin{array}{l}
 K_{2c}< 2\nu^2 {\rm ~~for ~even~ 1/\nu}, \vspace{5pt} \\
 K_{2c}< 3\nu^2 {\rm ~~for ~odd~ 1/\nu}.
\end{array}
\right.
\end{align}
At the commensurate fillings, the upper subband contributes a fractional conductance depending on the filling factor $\nu$ and an open channel in the half-gapped lower subband gives $e^2/h$. In consequence, upon decreasing $\nu$, the dip value decreases towards unity.
The appearance of the dips rely on strong electron-electron interactions, parametrized by $K_{2c}$. In principle, the smaller the filling factor is, the higher-order backscattering and thus the stronger interaction (smaller $K_{2c}$) is required for the occurrence of the corresponding dip. 
 Since the interaction parameter $K_{2c}$ itself also decreases when the electron density is reduced, as has been observed and analyzed in Refs.~\cite{Sato:2019,Hsu:2019}, we expect that the interactions in typical devices are sufficiently strong for multiple dips to be observable; see also Appendix~\ref{Appendix:Kjc} for the estimation on the $K_{2c}$ value. As shown in Fig.~\ref{Fig:conductance}, the dips appear around the edge of the second plateau, within the range of $\mu/E_g \in\{1, 4/3\}$. They become denser as the wire becomes more depleted.
Crucially, as long as electron-electron interactions are sufficiently strong, the positions and the values of the conductance dips are insensitive to specific material parameters---hence they are {\it universal}.

We remark that, in deriving Eq.~\eqref{Eq:conductance}, we have taken into account explicitly the effect of Fermi-liquid leads (characterized by $K_{c}^{\rm L}=K_{s}^{\rm L}=1$), as in early works on single-subband wires in the absence of backscatterings in the wire~\cite{Maslov:1995,Ponomarenko:1995,Safi:1995}. While electron-electron interactions are required for the appearance of the dips, the conductance value itself does not depend on the interactions in the wire. This result is similar to the conclusion of Refs.~\cite{Maslov:1995,Ponomarenko:1995,Safi:1995}, in the sense that the conductance is determined by the interaction strength in the leads instead of the wire.  
The interaction criterion summarized in Eq.~\eqref{Eq:conductance} reveals distinct scaling behaviors for the backscatterings at even- and odd-denominator fillings. Next, we fix the filling factor $\nu$, and discuss the higher-order backscattering mechanism separately for even and odd denominators.

\section{Even-denominator filling~\label{Sec:Even}}
\begin{figure}[t]
\centering
\includegraphics[width=0.72\linewidth]{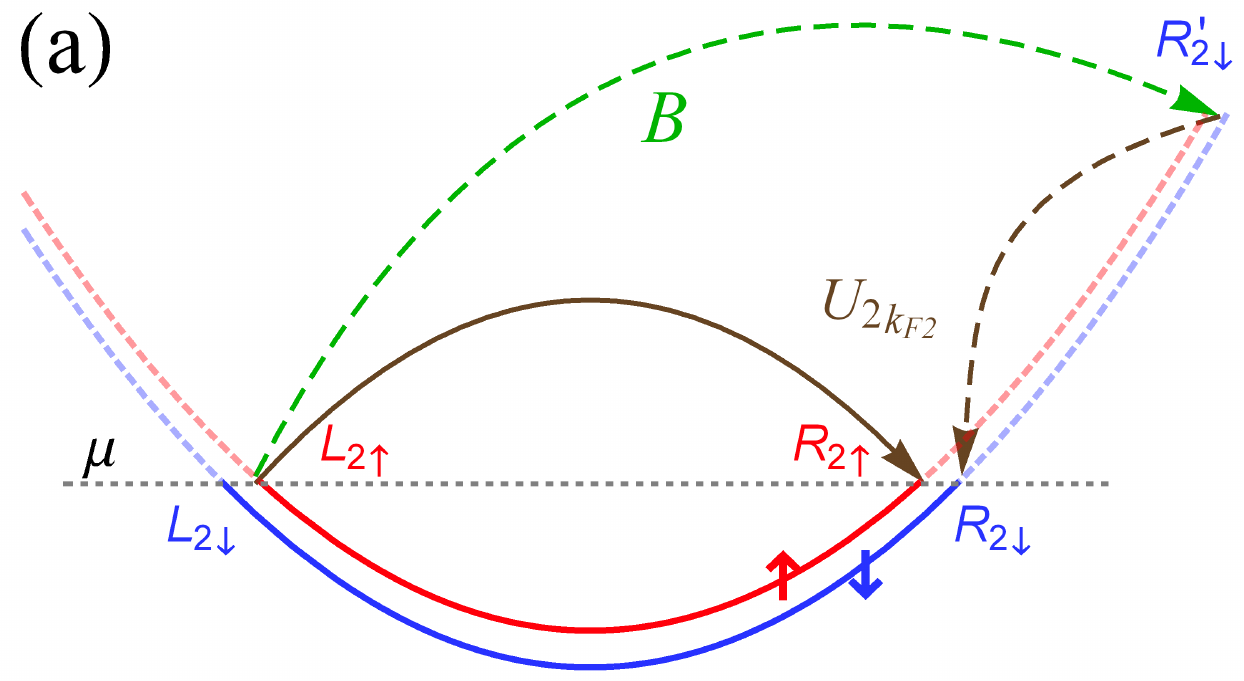}
\includegraphics[width=0.27\linewidth]{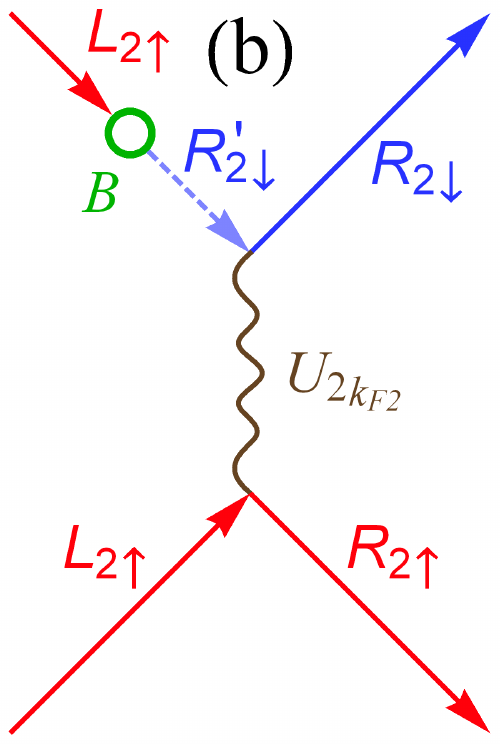}
\caption{Scattering process in the upper subband for $\nu=1/2$ (that is, $\kFo =2 \kFt$); for clarity, the two spin branches are separated and plotted in different colors. The lower subband, identical to the one plotted in Fig.~\ref{Fig:Setup}, is not shown. (a) Scattering process $O^{(2)}_{+}$.
The $\Hov$ term brings an electron from $\Ltu (-\kFt)$ to the intermediate state $\Rtd' (3 \kFt)$ (green dashed arrow). Electron-electron interactions allow $\Rtd'$ to forward scatter to $\Rtd (\kFt)$ (brown dashed arrow). The momentum difference is compensated by a backscattering process $\Ltu \to \Rtu$ (brown solid arrow).
(b) Diagram for $O^{(2)}_{+}$ scattering process in Panel (a). The red (blue) arrows indicate up-(down-)spin fermion fields. The wavy line and green circle mark Coulomb interaction and spin-flip backscattering by the helical magnetic field, respectively.
}
\label{Fig:g2}
\end{figure}

When the chemical potential is at the even commensurability $\nu = 1/(2n)$, it allows for the $(2n)$th-order helical-field-assisted scattering in the upper subband. 
In the RG framework, we keep the most relevant scattering, which consists of two terms, denoted as 
\begin{align} \label{Eq:O2n}
O^{(2n)}_{\sigma} =& \frac{ g^{(2n)} }{2} \big( \Rtd \da \Ltu \big) \big(  R_{2\sigma} \da L_{2\sigma}  \big)^{n}  \big( R_{2 \bar{\sigma}} \da L_{2 \bar{\sigma}} \big)^{n-1} + \hc,
\end{align}
with $g^{(2n)} \propto \Bov (U_{ 2k_{F2}})^{2n-1}$, the Fourier component of the Coulomb potential $U$, and $\sigma = -\bar{\sigma}  \in \{ \uparrow \; \equiv +, \downarrow \; \equiv - \}$. Figure~\ref{Fig:g2} illustrates the scattering process for $n=1$ and $\sigma=+$.
The helical field in $\Hov$ brings an electron from $\Ltu$ at the Fermi point $-\kFt$ to the intermediate state $\Rtd'$ at $3 \kFt$, which subsequently forward scatters to $\Rtd$ at $\kFt$ by electron-electron interactions. The momentum difference is compensated by a backscattering process, which brings an electron from $\Ltu$ at the Fermi point $-\kFt$ to $\Rtu$ at $\kFt$. One can generalize the diagram to describe scattering processes for $\sigma= -$, as well as for general $n$.
While the scattering involves states away from the Fermi level, the corresponding operator $O^{(2n)}_{\sigma} $ is written in terms of $R_{2\sigma}$ and $L_{2\sigma}$ defined at the Fermi level, where the bosonization and {\TLL} approach are valid.
Clearly, the initial value of $g^{(2n)}$ depends on the energies of the intermediate states and therefore the system details, such as the confinement potential and the curvature of the dispersion relation. 
Nevertheless, since the relevance of the operator $O^{(2n)}_{\sigma} $ is determined by its scaling dimensions instead of the bare coupling, we do not specify $g^{(2n)}$ here.

With the bosonization technique, we derive the RG flow equation for $\tilde{g}^{(2n)} =  g^{(2n)} a^2 /(\hbar v_{F2})$ with the Fermi velocity $v_{F2}$ of the upper subband,
\begin{align} \label{Eq:RG_even}
\frac{d \tilde{g}^{(2n)} }{d l } =& \big( 1 - 2 n^2 K_{2c} \big) \tilde{g}^{(2n)},
\end{align}
which indicates that the operator $O^{(2n)}_{\pm}$ is RG relevant for $ K_{2c}< 1/ (2 n^2)$. 
Before proceeding, we comment on other perturbations that could be relevant. First, the presence of a periodic potential (for instance, arising from the lattice) could enable umklapp scattering processes, in which electrons back scatter with the help of the crystal momentum~\cite{Giamarchi:2003}. However, the umklapp scattering is feasible only when the Fermi wave vector is on the order of the inverse lattice spacing, which is beyond the parameter regime of interest.
Therefore, we do not include the effect of the lattice periodicity here. 
Second, impurity scattering induced by disorder is in general RG relevant~\cite{Giamarchi:2003}, where the corresponding operator is given by $O_{\rm imp} = R_{2\sigma} \da L_{2\sigma} + \hc $ (other operators such as $R_{1\sigma} \da L_{2\sigma}$ do not commute with the most relevant $\Oov$ and therefore cannot be ordered). However, since $O_{\rm imp}$ has a different scaling dimension than $O^{(2n)}_{\sigma}$, we find that $\tilde{g}^{(2n)}$ grows faster than the coupling of $O_{\rm imp}$ under the RG flow for $K_{2c} < 1/(4n^2 -1)$. Moreover, when taking into account the RG flow up to the second order in couplings, the $K_{2c}$ value is also renormalized and reduced from its initial value, in favor of the dominance of $O^{(2n)}_{\sigma}$ over $O_{\rm imp}$. In consequence, the helical-field-assisted scattering dominates over the impurity scattering for sufficiently strong electron-electron interactions. 

To proceed, we express the operator as $O^{(2n)}_{\pm} = g^{(2n)} \cos \Phi_{\pm}^{e}$ in terms of the boson fields $\Phi_{\pm}^{e}$; see \app. 
Importantly, we show that $O^{(2n)}_{\pm}$ commute with each other and with $\Oov$, so that the three operators can be ordered simultaneously.
In wires with sufficiently strong interactions, meaning sufficiently small $K_{2c}$, the backscattering terms in Eq.~\eqref{Eq:O2n} open a gap in the upper subband, leading to dips in the conductance.

\begin{figure}[t]
\centering
\includegraphics[width=0.68\linewidth]{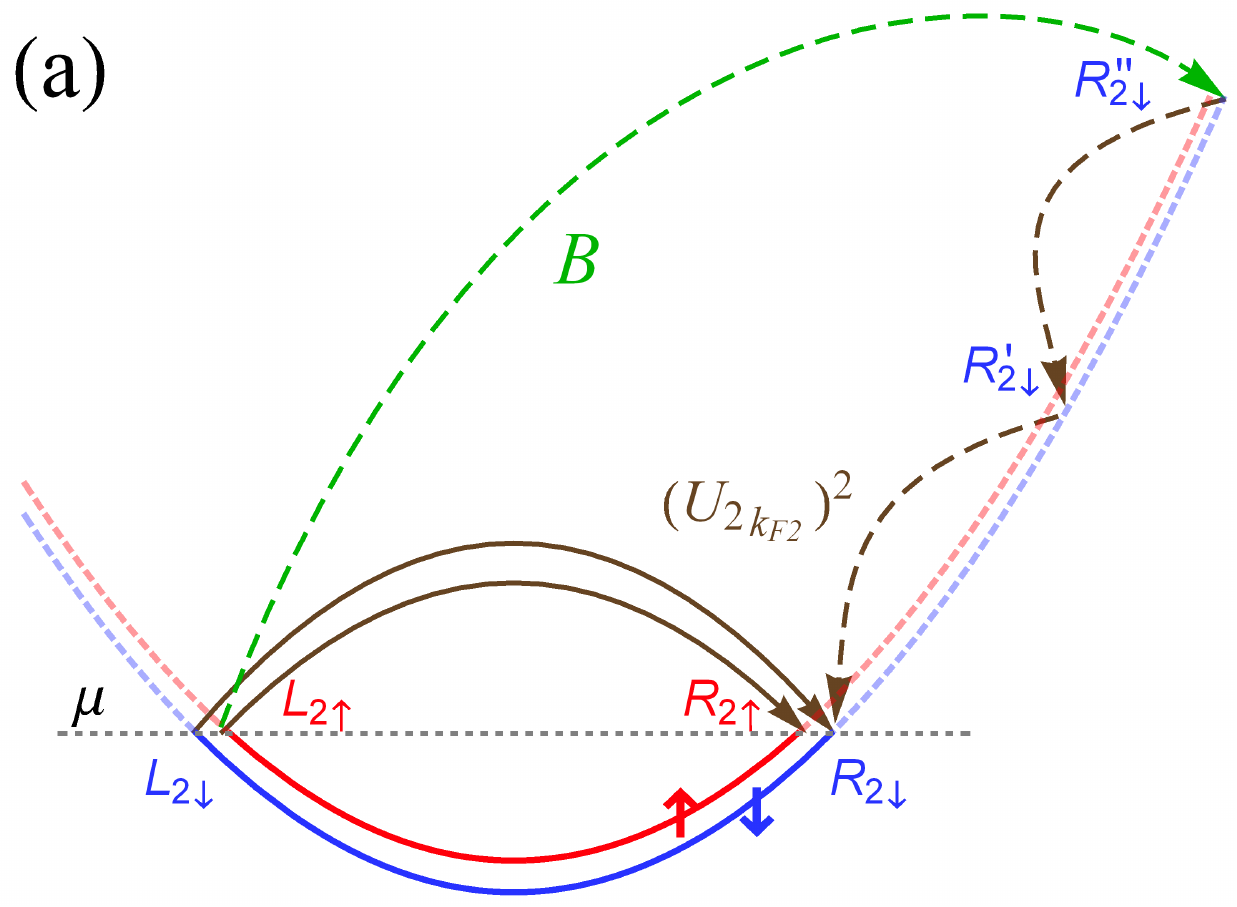}
\includegraphics[width=0.30\linewidth]{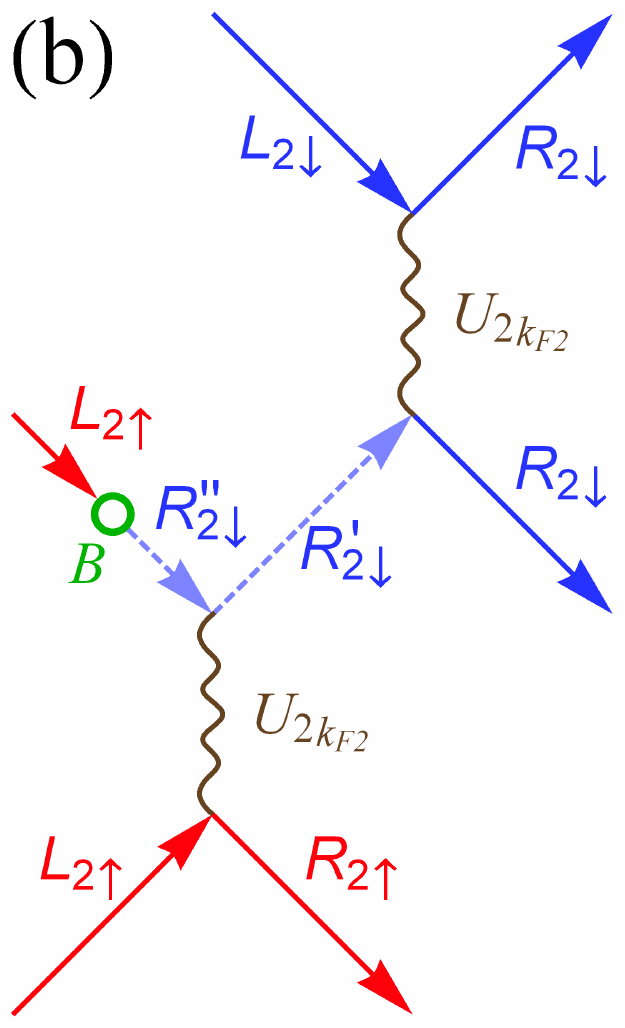}
\caption{The same as in Fig.~\ref{Fig:g2} but for an odd filling $\nu=1/3$ ($\kFo =3 \kFt$). (a) Scattering process $O^{(3)}$. 
The $\Hov$ term brings an electron from $\Ltu (-\kFt)$ to $\Rtd'' (5 \kFt)$ (green dashed arrow). With the help of electron-electron interactions, the electron scatters from $\Rtd''$, through $\Rtd' (3 \kFt)$, to $\Rtd (\kFt)$ (brown dashed arrows), accompanied by the backscattering processes $\Ltu \rightarrow \Rtu$ and $\Ltd \rightarrow \Rtd$ at the Fermi points (brown solid arrows).
(b) Diagram for $O^{(3)}$ scattering in Panel (a). The labels are the same as those given in Fig.~\ref{Fig:g2}.
}
\label{Fig:g3} 
\end{figure}

\section{Odd-denominator filling~\label{Sec:Odd}}
Now we turn to the odd case $\nu=1/(2n+1)$ and investigate the $(2n+1)$th-order helical-field-assisted scattering. In contrast to the even case, here the most RG relevant process consists of a single term, 
\begin{align}
O^{(2n+1)} = \frac{g^{(2n+1)}}{2}  \big( \Rtd\da \Ltu \big) \big( \Rtu\da \Ltu \big)^{n}  \big( \Rtd\da \Ltd \big)^{n} + \hc,
\end{align}
with $g^{(2n+1)} \propto \Bov (U_{ 2k_{F2}})^{2n}$; Figure~\ref{Fig:g3} illustrates the $O^{(3)}$ process, which is similar to Fig.~\ref{Fig:g2} but involves higher-order scattering processes. 

Upon bosonization, we derive the RG flow equation,
\begin{align} \label{Eq:RG_odd}
\frac{d \tilde{g}^{(2n+1)} }{d l} =& \frac{3  - (2n+1)^2 K_{2c} }{2}\, \tilde{g}^{(2n+1)},
\end{align}
with $\tilde{g}^{(2n+1)} =  g^{(2n+1)} a^2 /(\hbar v_{F2})$. The operator $O^{(2n+1)}$ is RG relevant for $ K_{2c}< 3/ (2 n+1)^2$. 
In addition, we find that the process $O^{(2n+1)}$ dominates over the impurity scattering $O_{\rm imp}$ for sufficiently strong interactions $K_{2c} < 2/ [(2n+1)^2 -1]$. 
Again, the reduced $K_{2c}$ value under the (second-order) RG flow tends to enhance the dominance of $O^{(2n+1)}$ over $O_{\rm imp}$, so we have dominant helical-field-assisted scattering in a strongly interacting system.

The operator can be written as $O^{(2n+1)} =  g^{(2n+1)} \cos ( 2 \sqrt{2n+1} \Phi_{+}^{o} )$ with the transformation given in Eq.~\eqref{Eq:T_odd}; see \app.
In the presence of strong interactions, the $O^{(2n+1)}$ term leads to a partial gap in the upper subband. 
As a result, the system contains a fractional (helical) {\TLL} in the upper (lower) subband.
The conductance, with contributions from both subbands, is summarized in Eq.~\eqref{Eq:conductance}. 
In the bosonic language, when $O^{(2n+1)}$ is ordered, the $\Phi_{+}^{o}$ field is pinned at multiples of $\pi/\sqrt{2n+1}$, while the excitations correspond to kinks in $\Phi_{+}^{o}$, where the field changes its value between neighboring minima. 
We find that such a kink carries a charge of $q_{\rm k} = \nu e$ and zero spin.
This fractional charge can be examined through shot noise, similar to the proposed setup in Ref.~\cite{Cornfeld:2015}.
Here, $q_{\rm k}$ reveals the hierarchy of the Laughlin states~\cite{Laughlin:1983}, similar to an array of one-subband quantum wires~\cite{Kane:2002,Klinovaja:2014d,Meng:2014c,Klinovaja:2015b,Santos:2015} 
and Rashba wires~\cite{Oreg:2014,Aseev:2018,Shavit:2019} in magnetic fields.

\section{Discussion~\label{Sec:Discussion}}
In this work, we demonstrate a mechanism for universal conductance dips and fractional excitations. While the mechanism can take place in a wide variety of materials and setups, we expect that the most natural realization is to utilize III-V semiconducting quantum wires, where nuclear spins couple to conduction electrons through the hyperfine interaction. The wires realize Kondo lattices, in which nuclear spins are ordered into a helical pattern at dilution fridge temperature [e.g., $O(10~{\rm mK})$--$O(100~{\rm mK})$ for GaAs and InAs wires], as predicted theoretically in Refs.~\cite{Braunecker:2009a,Braunecker:2009b} and indicated in the experiment with cleaved edge overgrowth GaAs quantum wires in Ref.~\cite{Scheller:2014}.
The ordered nuclear spins generate an internal helical magnetic field required for our mechanism. 
In general terms, this system consists of a Kondo lattice in which the couplings between itinerant charge carriers and localized spins are weak so that the RKKY coupling dominates over the Kondo screening~\cite{Tsvelik:2019a,Yevtushenko:2020}.
With the tunable dimensionality of our setup, we demonstrate that Coulomb interaction between the charge carriers can trigger the formation of strongly correlated fermions even in this weak-coupling limit. 
Remarkably, this realization relies on ingredients naturally present in semiconductor wires and can host fractional excitations, resembling fractional quantum Hall states even without external magnetic fields.

Alternatively, magnetic dopants (e.g., Mn) can substitute nuclear spins in creating the RKKY-induced helical field. Yet differently, the quasi-one-dimensional Kondo lattice can be fabricated out of heavy-fermion compounds, where itinerant electrons and localized spins interact through exchange couplings~\cite{Prochaska:2020}.
In these alternative systems, both the transition temperature and the helical field strength can be enhanced due to larger exchange couplings~\cite{Klinovaja:2013}. 
Since the helix is susceptible to magnetic fields~\cite{Stano:2014} and elevated temperatures~\cite{Aseev:2017,Aseev:2018}, field- and temperature-dependent conductance might be used to seek additional signatures, such as disappearance of the dips and restoration of the standard plateaus. 
The observation of universal conductance dips would indicate the helix-assisted higher-order scattering and therefore the presence of the helix itself, supplementary to uniformly dropped plateaus~\cite{Scheller:2014}.
Apart from the above Kondo-lattice scheme, the helical magnetic field can be artificially generated by the external magnetic field in Rashba wires or by depositing nanomagnets~\cite{Klinovaja:2012b}, which are target systems of active researches on  topological matters in the quasi-one dimension.
 
For the purpose of observation, as disorder in quantum wires can mask the conductance features, the predicted dips should be searched for in wires in the ballistic regime. 
Since the proposed scatterings involve intermediate states at relatively high energies, quantum wires capable of sustaining these intermediate states are desirable, which can be achieved using, for instance, quantum wires that confine a few more transverse modes~\cite{Patlatiuk:2020}.
To distinguish the predicted dips from Fabry-Perot oscillations in a finite-size wire, one might suppress the latter features through device design, such as wire length or smoothness of the contact between the wire and leads~\cite{Aseev:2017,Aseev:2018}. 
In addition, while the Fabry-Perot oscillations, if present, occur on both of the first and second conductance plateaus, the dips appear only on the second plateau and are inevitably accompanied by nonstandard quantized values of the two plateaus in a certain temperature range. 
Since the conductance dips can be observed in straightforward transport measurements, our prediction can be verified through systematic investigations on the temperature-dependent conductance upon varying carrier density and/or wire geometry.

\begin{acknowledgments}
We thank T.-M.~Chen, S.~Matsuo, D.~Miserev, T.~Patlatiuk, S.~Tarucha, H.~Weldeyesus, and D.~M.~Zumb{\" u}hl for interesting discussions.
This work was financially supported by the JSPS Kakenhi Grant No.~16H02204 and No.~19H05610, by the Swiss National Science Foundation (Switzerland), and by the NCCR QSIT. This project received funding from the European Unions Horizon 2020 research and innovation program (ERC Starting Grant, Grant agreement No.~757725). 
\end{acknowledgments}

\appendix
\section{DETAILS OF THE CALCULATION~\label{Appendix:Details}}
In this section we present the details of our calculation.
To analyze the higher-order backscattering processes, we employ the bosonization technique~\cite{Kane:2002,Giamarchi:2003,Meng:2013,Klinovaja:2014a,Klinovaja:2014d,Klinovaja:2015,Klinovaja:2015b}. We express the slowly varying fields [see Eq.~\eqref{Eq:SlowVarying}] as 
\begin{align}
\Rjs (x) =& \frac{1}{\sqrt{2 \pi a}} e^{i \phi_{Rj\sigma} (x) }, ~ \Ljs (x) =  \frac{1}{\sqrt{2 \pi a}}  e^{i \phi_{Lj\sigma} (x) },
\label{Eq:bosonization}
\end{align}
with the chiral bosonic fields $\phi_{\ell j\sigma}$. In the above, we omit the Klein factors, and use the indices $\ell \in \{ R \equiv +, L \equiv -\}$, $j \in \{1,2\}$, and $\sigma \in \{ \uparrow \, \equiv + ,\,  \downarrow \, \equiv - \}$ to label the chirality, subband, and spin, respectively.

For the odd-denominator filling $\nu = 1 / (2n+1)$, we impose the following commutation relation
\begin{align}
[\phi_{\ell j\sigma} (x), \phi_{\ell' j'\sigma'} (x')] =& i \pi \ell  \delta_{\ell \ell'} \delta_{jj'} \delta_{\sigma \sigma'} {\rm sign}(x-x').
\label{Eq:commutator1} 
\end{align}
Next, we express the operator $O^{(2n+1)}$ as a cosine~\cite{Kane:2002,Klinovaja:2014a,Klinovaja:2014d,Klinovaja:2015,Klinovaja:2015b},
\be
O^{(2n+1)} &=& g^{(2n+1)} \cos \big( 2 \sqrt{2n+1} \Phi_{+}^{o} \big),
\ee
through the transformation $\Psi^{o} = T^{o} \psi $ with $\Psi^{o} \equiv (\Phi_{+}^{o}, \Phi_{-}^{o}, \Theta_{+}^{o}, \Theta_{-}^{o})^t$, $\psi \equiv (\phiRtu, \phiRtd, \phiLtu,  \phiLtd )^t$, the transpose operator $t$, and the matrix $T^{o}$ given by 
\be 
T^{o} &=& \frac{1}{2 \sqrt{2 n +1} } \left(
\begin{array}{cccc}
 - n &   - (n+1) & n+1 &  n  \\
 -(n+1)  & - n & n & n+1 \\
-n & n+1 & n+1 & -n \\
n+1 & -n & -n & n+1
\end{array}
\right) . \nonumber \\
\label{Eq:T_odd} 
\ee
The new fields satisfy the commutation relations 
\begin{subequations}
\label{Eq:commutator2}
\begin{eqnarray}
&& [\Phi_{\pm}^{o} (x), \Theta_{\pm}^{o} (x')] = \frac{i \pi}{2} {\rm sign}(x'-x), \\
&& [\Phi_{\pm}^{o} (x), \Theta_{\mp}^{o} (x')] = 0, \\
&& [\Phi_{+}^{o} (x), \Phi_{-}^{o} (x')] = [\Theta_{+}^{o} (x), \Theta_{-}^{o} (x')] = 0.
\end{eqnarray}
\end{subequations}
To evaluate the charge of the excitations, we express the charge density operator as~\cite{Giamarchi:2003}
\begin{align}
\rho_{\rm c} (x) =& \frac{e}{2\pi} \sum_{\ell j \sigma} \partial_{x} \phi_{\ell j \sigma} (x) .
\end{align}
Using Eq.~\eqref{Eq:T_odd} and performing the integral over space, we evaluate the charge
\begin{align}
q_{\rm k} =& \int_{\rm kink} dx \; \rho_{\rm c} (x) = \frac{e}{\pi \sqrt{2n+1}} \Delta \Phi_{+}^{o} ,
\end{align}
with the change $\Delta \Phi_{+}^{o} $ of the field across a kink. 
Therefore, a kink where the $\Phi_{+}^{o}$ field changes its value by $\pi/\sqrt{2n+1}$ carries a charge of $e / (2n+1)$.
In addition, the linear-response conductance from the fractional {\TLL} can be computed straightforwardly as in Refs.~\cite{Oreg:2014,Meng:2014d}. The results are given in the main text. 

For even-denominator fillings $\nu = 1 / (2n)$, we use the commutator in Eq.~\eqref{Eq:commutator1} for $j=1$, and the following generalized commutation relation for $j=2$, 
\begin{align}
[\phi_{\ell 2\sigma} (x), \phi_{\ell' 2\sigma'} (x')] =& i \pi M_{\ell \sigma \ell' \sigma'} {\rm sign}(x-x'),
\label{Eq:commutator}
\end{align}
with $M_{\ell \sigma \ell' \sigma'}$ being an integer depending on the chirality and the spin. 
In addition, the chiral boson fields for $j=1$ and $j=2$ commute with each other.
Defining the new index,
\be
 p,p' \in \{ 1 \equiv R\uparrow, \; 2 \equiv R\downarrow, \; 3 \equiv L\uparrow, \; 4 \equiv L\downarrow \},
\ee
we can write $M_{pp'}$ in the following matrix form, 
\be
\left(
\begin{array}{cccc}
1 & -(2n-1) & -1 & 0  \\
-(2n-1)  & 1 & 0 & 1 \\
-1 & 0 & -1 & 2n-1 \\
0 & 1 & 2n-1 & -1
\end{array}
\right).
\label{Eq:M}
\ee
Denoting $\Psi^{e} \equiv (\Phi_{+}^{e}, \Phi_{-}^{e}, \Theta_{+}^{e}, \Theta_{-}^{e})^t$, 
we make the transformation $\Psi^{e} = T^{e} \psi $, with the matrix $T^{e}$ given by 
\be
T^{e} =
\left(
\begin{array}{cccc}
- n & - n & n+1 &   n-1  \\
n-1  & n+1 & - n &  - n \\
-f_n & -(2n^2-1) f_n & 2n(n-1) f_n & 0 \\
0 & -2n(n-1) f_n & (2n^2-1) f_n & f_n
\end{array}
\right), \nonumber \\
\label{Eq:T_even}
\ee
with $f_n \equiv 1/[4n(4n^2-4n-1)]$.
It can be shown that the new fields $\Phi_{\pm}^{e}$ and $\Theta_{\pm}^{e}$ satisfy Eq.\eqref{Eq:commutator2} upon replacing the superscript $o \to e$. With Eqs.~\eqref{Eq:commutator}--\eqref{Eq:T_even}, the operators $O^{(2n)}_{\pm}$ can be expressed in terms of $\Phi_{\pm}^{e}$, as given in the main text. Similarly to the odd case, the charge at the domain wall in the $\Phi_{\pm}^{e}$ fields is given by $e/(2n)$.

Since the $\Phi_{\pm}^{e}$ fields arise from the unconventional commutation relations in Eq.~\eqref{Eq:commutator}, it is not obvious  whether a partial or full gap is opened in the upper subband. We therefore employ a Green's function-based approach~\cite{Maslov:1995,Meng:2014d} and compute directly the linear-response conductance. By modeling the system as a fractional Tomonaga-Luttinger liquid wire and Fermi-liquid leads, we find the values for the conductance dips, as presented in the main text.

\begin{figure}[t]
\centering 
 \includegraphics[width=\linewidth]{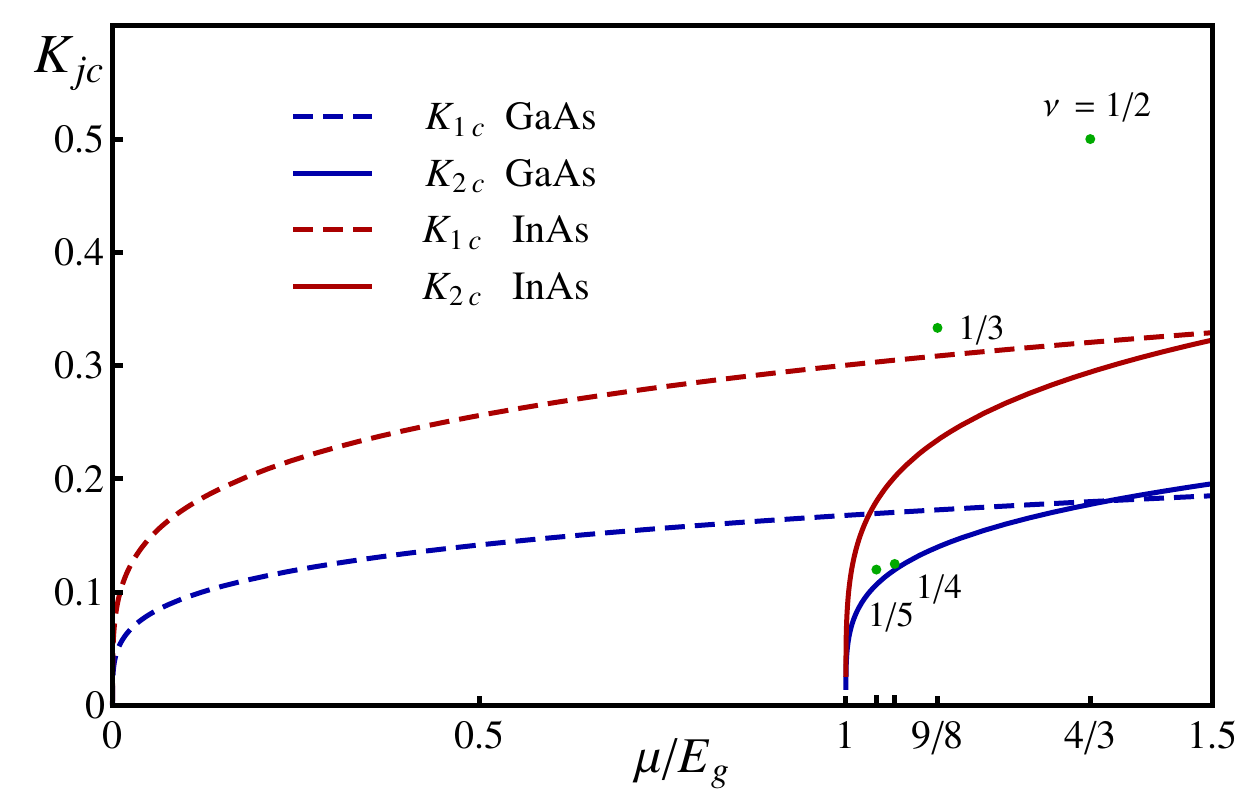}
 \caption{Interaction parameters $K_{jc}$ for subband $j \in \{1,2\}$ in a GaAs (blue) and InAs (red) quantum wire as functions of $\mu/E_g$. 
The green dots mark the $K_{2c}$ conditions for $\nu = 1/2, \cdots, 1/5$ at the corresponding positions [see Eq.~\eqref{Eq:conductance} in the main text]. 
The adopted parameter values are $\epsilon=12.9~\epsilon_{0}$ ($15.15~\epsilon_{0}$), $m=0.067~m_{e}$ ($0.023~m_{e}$) for GaAs (InAs), $D=400~\mu$m, $w=100$~nm, and $d=10~$nm. }
\label{Fig:Kjc}
\end{figure}

\section{ESTIMATION FOR THE INTERACTION PARAMETER~\label{Appendix:Kjc}}
In this section we estimate the interaction parameter $K_{jc}$ for subband $j$. By generalizing the formula in Refs.~\cite{Giamarchi:2003,Hsu:2018b,Sato:2019} for wires with multiple subbands, we obtain
\begin{equation}
K_{jc} = \left[1+\frac{2e^2}{\pi ^2 \epsilon \hbar v_{Fj}} \ln \left| \frac{D}{ {\rm max} (w_j, d) }\right| \right]^{-1/2},
\label{Eq:Kjc}
\end{equation}
with the dielectric constant $\epsilon$, the Fermi velocity $v_{Fj} = \hbar k_{Fj}/m$, the Fermi wave vector $k_{Fj}$, the effective mass $m$, the screening length $D$ (the distance between the wire and a nearby metallic gate), and the thickness $d$ of the wire. Defining the effective width $w_j = \sqrt{\langle y^2\rangle_j}$ (averaged with respect to the $j$-th subband electron wavefunction) and taking a simple harmonic confinement potential along $y$ direction with the subband spacing $E_g = \hbar^2/(2mw^2)$, we find $w_1 = w$ and $w_2=\sqrt{3} w$. 
While our backscattering mechanism does not rely on any specific confinement, here we adopt the parabolic confinement potential in order to make concrete estimation for $K_{jc}$; adopting other types of confinement would not make qualitative difference.

The estimated $K_{jc}$ values for GaAs and InAs wires are shown in Fig.~\ref{Fig:Kjc}, where the green dots mark the upper bounds for the $K_{2c}$ values required for our mechanism. To be specific, for the dips corresponding to $\nu = 1/2, ~1/3, ~1/4, ~1/5, ~\cdots$ to emerge, we need $K_{2c}$ below $0.5, ~0.33, ~0.13, ~0.12, ~\cdots$, respectively.
Since $v_{Fj}$ decreases when approaching the subband bottom, we find sufficiently low $K_{2c}$ values in the regime of interest, indicating sufficiently strong electron-electron interactions for the helical-field-assisted backscattering.
Specifically, for the adopted parameters we find sufficiently small $K_{2c}$ values for $\nu = 1/2, 1/3$ in InAs wires and for $\nu$ up to $1/5$ in GaAs wires. 
In conclusion, the $K_{2c}$ criterion for the higher-order scattering can be fulfilled in the realistic regime, allowing for multiple conductance dips appearing in typical semiconductor wires.

\end{document}